# A Pointillism Approach for Natural Language Processing of Social Media

Peiyou Song
Computer Science
University of New Mexico
Albuquerque NM 87131
Email: peiyou@cs.unm.edu

Anhei Shu
Computer Science
Rice University
Houston TX 77251

Anyu Zhou
Computer Science
Harbin Engineering University
Harbin 150086, China

Dan Wallach
Computer Science
Rice University
Houston TX 77251

Jedidiah R. Crandall
Computer Science
University of New Mexico
Albuquerque NM 87131

*Abstract*—
The Chinese language poses challenges for natural language processing based on the unit of a word even for formal uses of the Chinese language, social media only makes word segmentation in Chinese even more difficult. In this document we propose a pointillism approach to natural language processing. Rather than words that have individual meanings, the basic unit of a pointillism approach is trigrams of characters. These grams take on meaning in aggregate when they appear together in a way that is correlated over time.

Our results from three kinds of experiments show that when words and topics do have a meme-like trend, they can be reconstructed from only trigrams. For example, for 4-character idioms that appear at least 99 times in one day in our data, the unconstrained precision (that is, precision that allows for deviation from a lexicon when the result is just as correct as the lexicon version of the word or phrase) is 0.93. For longer words and phrases collected from Wiktionary, including neologisms, the unconstrained precision is 0.87. We consider these results to be very promising, because they suggest that it is feasible for a machine to reconstruct complex idioms, phrases, and neologisms with good precision without any notion of words. Thus the colorful and baroque uses of language that typify social media in challenging languages such as Chinese may in fact be accessible to machines.

## I. INTRODUCTION

Social media poses many challenges for natural language processing. Many of these challenges center around the concept of a word. For example, social media users might invent new words called neologisms, which can express more meaning than the original word or evade content filtering. An example in English would be "Intarweb" to refer to the Internet, by using the neologism "Intarweb" net users are adding the additional meaning that the Internet and web are not something that is fully understood.

Chinese social media compounds these problems that are caused by the notion of a word, both because the written Chinese language does not delimit words with spaces and because Chinese net users use neologisms very heavily. How can we track trends, discover memes, and perform other basic natural language processing techniques for Chinese social media when it is not even clear that the problem of segmenting Chinese social media into words is a tractable problem?

This document is centered around a thought experiment: how much can machines understand about trends in social media for challenging languages such as Chinese without any lexicon or notion of words? We propose a pointillism approach to natural language processing, and through experiments show that longer words and phrases can be put back together based only on temporal correlations of trigrams.

To motivate the pointillism approach to natural language processing in this document, we focus on the Chinese language. A unique feature of the Chinese writing system is that it is a linear sequence of non-spaced ideographic characters. The fact that there are no delimiters between words poses the well-known problem of segmentation. A natural language processing system with a lexicon could perform quite well, however, the unknown words which are not registered in the lexicon, become the bottle-neck in terms of precision and recall [1]. For Chinese, Chooi and Ling [2] observed that if one can obtain good recall for unknown words, the overall segmentation is better. However, in Chinese social media unknown words are used regularly.

In this document, we illustrate an approach to recognize trends, not based on words as are traditional methods but based on trigrams, and use the trends to find words which can be either known or unknown. This method uses only a corpus with a time series, meaning that a system dictionary and grammar knowledge is not required. The underlying concept of the proposed method is as follows. We regard the problem of trend analysis as finding the trigrams which have the same trend. We observe that the trigrams that have correlated trends over a long period of time have a high probability that they belong to the same topic or even belong to the same word. We concatenate the trigrams back until it becomes a word or a phrase.

The rest of this document is organized as follows. First, we give some background about the Chinese language and Sina Weibo in Section II. Then Section III discusses some preliminary observations that motivate the approach and Sections V and IV explain the key algorithms and procedures for discovering bigram and trigram trends and then concatenating trigrams to form a word or phrase. Our experimental methodology and results are explained in Section VI. Then the discussion in Section VII and related works in Section VIII are followed by the conclusion.

## II. BACKGROUND, IN BRIEF

English speakers expect words to be separated by whitespace or punctuation. In Chinese, however, words are simply concatenated together. Therefore, in order to understand Chinese text, the first thing that we need to do is to divide the sentences into word segments. The problem of mechanically segmenting Chinese text into its constituent words is a difficult problem. During the process of segmentation of Chinese, two main problems are encountered: segmentation ambiguities and unknown word occurrences.

Since social media is heavily centered around current events, it contains many new named entities that will not appear in even the most comprehensive lexicons. Neologisms, another type of unknown words, are created to express a new meaning or the same meaning with different nuance. Neologisms are also an integral part of social media.

For example: in the following post from a microblogging site of China, weibo.com, there is at least one unknown word in each sentence.

苦逼小青年-S：每天被这么多不认识的人@ 真的是有一种受宠若惊的感脚。但作为一个女丝我只能辜负大家对俺的厚爱了。对唔住啊！

SINA Weibo has the most active user community among the other top 3 portal sites: Tencent, Sohu and NetEase. Moreover, SINA has the best relationship with the Chinese government. Recently, even government employees and government media use SINA weibo to broadcast news and other events [6].

Microblogging entails real-time sharing of content that is specific to a time and audience. This is in contrast to traditional media that has a longer news cycle and a prolonged process that makes the content and timing of the content more uniform. Compared to other online corpora, microblogs are distinguished by short sentences and casual language. Most microblog sites limit the maximum length of a post to 140 UTF-8 characters[1], demanding precise and clear execution. Microblogs are important birthplaces of new words. Moreover, microblog posts have timestamps. This information is essential to our pointillism approach.

## III. OBSERVATIONS

To illustrate how a word can create ambiguities for word segmenters, we use 中华人民共和国 (People's Republic of China) as an example. This is neither a neologism nor an unknown word that cannot be found in the dictionary, but it is a good example of segmentation ambiguities and gram trends.

The word 中华人民共和国 (People's Republic of China) is seven characters long and has smaller words within, as shown in Figure 1.

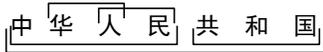

Fig. 1. People's Republic of China.

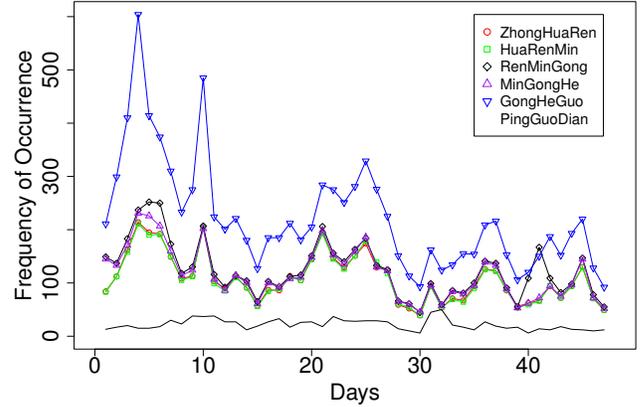

Fig. 2. Trigram trends for 中华人民共和国 on Weibo.

Figure 2 shows the plot of trigram frequency of occurrence for a period of 47 days[2]. 中华人民共和国 has 5 trigrams: 中华人 (ZhongHuaRen), 华人民(HuaRenMin), 人民共 (RenMinGong), 民共和 (MinGongHe), and 共和国 (GongHeGuo). What can be seen in Figure 2 is that trigrams from 中华人民共和国 have a very distinctive temporal correlation when compared to other trigrams, such as the trigram 苹果电 (PingGuoDian, the first three characters of 苹果电脑, or Apple Computers). The x-axis is time in days. The y-axis is the number of occurrences of the trigram per day for our dataset, which at the time this data was taken was only about 2% of all Weibo posts. What is most interesting are trigrams that are not words themselves, but through time correlations serve as a sort of glue to hold trigrams which belong to the same word together.

The observations that there are time correlations that appear among trigrams leads to the question: can we use this feature to find words, phrases, memes or even find topics? In this research, we examine the possibility of using only time correlation information of grams to concatenate words and phrases without a lexicon or knowledge of the grammar.

## IV. PROCEDURES

In this section, we explain the key algorithms and procedures for discovering bigram and trigram trends and then concatenating trigrams to form a word or phrase.

**Step 1: Collecting posts with time sequences.**

Starting from 23 July, 2011[3], we send a request every second to the public timeline and the Weibo server returns roughly 200 posts responding to each request. These 200 posts are not continuous in terms of post ID .

**Step 2: Count the frequency of occurrence of grams.**

Given a Chinese text with time series information, the system will divide the text into trunks of three consecutive characters that are called trigrams. Our system obtains the frequency of occurrence of each trigram hourly. In this document, we mostly use the frequency of occurrence of each trigram for a daily basis.

**Step 3: Check if the trigram is a valid root.**

For convenience, we call the program connecting the trigrams based on time correlation information "Connector." We call the first trigram we feed into Connector the *root trigram*. Not all trigrams can be used as the root for concatenating trigrams. There are two kinds of trigrams that cannot be root trigrams.

First: trigrams that are rarely used where most of their daily frequencies are zeros. This is important because if a vector has too many zeros this would bias the cosine similarity value.

Second: trigrams which have no obvious fluctuation in our examination period. We measure this by obtaining the cosine similarity value between the trigram and the constant vector $\{1, 1, ..., 1\}$. If the value is higher than a certain threshold, 0.98 in our current system, then it is considered too normal and will be treated as an invalid root for concatenation.

**Step 4: Find time correlation of grams.**

The trigrams that have temporal correlations and overlap will be concatenated together. The detailed algorithms for this step are described in the next section.

## V. Algorithms

Cosine similarity is used to judge whether two trigrams have correlated trends.

$$cos.Sim = \frac{<A_i, B_i>}{\sqrt{\sum_{i=1}^{n} {A_i}^2} \times \sqrt{\sum_{i=1}^{n} {B_i}^2}}$$

where $<,>$ denotes an inner product between two vectors. We tested three types of vectors, the daily frequency of trigrams (FT), the daily difference in frequency of trigrams (DFT) and the daily rate of change in frequency (CFT).

$$CFT_{day}(trigram) = \log_{10} \frac{FT_{day}(trigram)}{FT_{day-1}(trigram)}$$

$$DFT_{day}(trigram) = FT_{day}(trigram) - FT_{day-1}(trigram)$$

During our initial experiments we found that, for some trigrams, CFT cosine similarity works better, and for other trigrams DFT works better. In light of this, we used an SVM (Support Vector Machine) to learn which parameters we should use for connecting the words. The features we used for the SVM were cosine similarity between the *root trigram* and the constant vector $\{1, 1, ..., 1\}$, maximum change rate of the root, minimum FT, maximum FT, mean of FT, and the length of the time period of vectors. The corresponding variable $y$ is a vector of FT, DFT and CFT. We used the libsvm package [9] to train our model. We labeled 100 data points for training by hand, and used this model to decide which vector should be used in finding the relationship between trigrams. Once we decide which vector we are going to use to find similarity, we concatenate the trigrams.

Let us use an English meme, "putting lipstick on a pig", as an example to explain how we concatenate trigrams.

Suppose Connector has thus far produced "putting lipstick on a". The system now considers which trigram from the set {"on a boat", "on a pig", "on a clear" and "on a plain"} it should concatenate next. The *root trigram* is the first trigram, "putting lipstick on a", with which Connector starts. The *parent trigram* is the last trigram appended to the phrase (at this point "lipstick on a" is the parent trigram). The *stem trigram* is the one which has the lowest frequency in "putting lipstick on a". We use the median value of the frequency vector to decide which trigram's frequency is the lowest, and the *stem trigram* should be closest to the real frequency of the phrase. To decide whether we should connect a trigram to the phrase we are working on, we compare the cosine similarity value between each candidate trigram and the *root* (*simRoot*), *parent* (*simParent*), and *stem trigram* (*simStem*). We use simScore to measure how close a gram is to the phrase.

$$simScore = max\{simRoot, simParent, simStem\}$$

The candidate trigrams are sorted by simScore. We recursively search for the next trigram to concatenate by a depth-first traversal into highest-score nodes first. Only the top 5 candidate trigrams are considered. The traversal stops when all candidate trigrams have a simScore lower than a threshold, which is set manually to 0.97. We set a threshold of 60 seconds for Connector to concatenate trigrams, to avoid unbounded searches.

$$simPath = \prod_{i=1}^{n-2} \{cos.Sim(simStem, trigram_i)\}$$

where, $n$ is the number of characters in the phrase.

The phrase which has the highest simPath is considered to be the final result. If there are other phrases which also have a *simPath* higher than the threshold, we output them as possible results, as shown in Section VII-A.

## VI. Evaluation

To gain a better understanding of the capabilities and limitations of connecting trigrams into longer words and phrases, we performed three experiments. For these experiments, we were interested in both known words (in the dictionary) and unknown words or phrases (not in the dictionary). Unknown words are the already existing words that are not in the dictionary. They include named entities and neologisms. Named entities can be the names of people, companies, movies, and anything that is given a name. Neologisms are words created recently, they are usually not words before they are created.

We collect three different types of word and phrase sets: long words or phrases from Wiktionary, hot keywords or phrases from weibo.com, and 4-character idioms from Wiktionary. Words or phrases in Wiktionary include both known

| | LCP | UP |
|---|---|---|
| Named entities (133) | 0.5 | 0.79 |
| New words (16) | 0.53 | 0.93 |
| Phrases (244) | 0.44 | 0.75 |
| 4-character idioms (29) | 0.90 | 0.90 |
| Total (422) | 0.50 | 0.79 |

and unknown words and they usually do not have an obvious trend[4]. 4-character idioms are known words and do not usually have obvious trends. Hot keyword lists from weibo.com are newly created words or phrases, and have obvious trend(s) in a certain period.

The procedures of these experiments start from Step 3 in Section IV (Step 1 and 2 have been done before the experiments). The output of Connector is compared with the original dataset, and judged by a human to determine if it is a correct result.

For this document, because our emphasis is on connecting trigrams into longer words and phrases rather than on how to detect trends, we assume that the word or phrase exists in our database and that the root trigram is given. For this reason we have no negative observations, only positive observations. This is why we focus on precision and do not report recall in this document. To compare two scenarios, one where Connector must only match words or phrases in a lexicon and one where a human can judge if Connector constructed a valid word or phrase, we report two different precision scores: lexicon-constrained and unconstrained.

Lexicon-constrained precision (LCP) is calculated as:

$$LCP = \frac{\{Matching\ Wiktionary\ or\ Weibo\}}{Valid\ Root\ Trigrams}$$

Unconstrained precision (UP) is calculated as:

$$UP = \frac{\{Matching\ Wiktionary\ or\ Weibo\} \cap \{Correct\}}{Valid\ Root\ Trigrams}$$

$\{Matching\ Wiktionary\ or\ Weibo\}$ are those exactly match with the words or phrases listed by them. $\{Correct\}$ are those different from the phrases listed by the source, but judged to be correct by a human.

### A. Long words and phrases from Wiktionary

We collected 500 words or phrases which were longer than 3 characters from wiktionary.com [10]. The results for each type are listed in Table I.

Except for 4-character idioms, the LCP scores are less than 0.55. LCP needs to be put in the proper context for our system. For example, 一房一厅 (One bedroom and one living room) becomes 一房一妻制 (One house one wife policy). The latter is a newly created phrase describing a new social phenomenon.

---

[4]A trigram having an obvious trend means it has a deviation from its standard tendency. For example a trigram involved in a news event or a hot topic.

We also found that the results tended to be effected by new events. For example, for one named entity, 中国银行业监督管理委员会 (China Banking Regulatory Commission) in wiktionary, the result of Connector is 中国银行百年行庆 (Bank of China One Hundred Year Anniversary).

The fact that UP scores tend to be much higher than LCP scores in or results implies that even with a relatively comprehensive lexicon, such as Wiktionary, the lexicon is incomplete and not up to date. This motivates the pointillism approach: our aim in this document is to move away from lexicons. Using an online corpus and analysing time enables us to extract newly created words/phrases, and even topics, with unknown words in real time.

Moreover, we found that when there are several candidate phrases which start from the same root trigrams, Connector would return the one which is used most frequently in a certain period of time, and this one may not be the one listed on Wiktionary. If we shorten the experiment duration and just include the period which had a trend, Connector would just return the exact phrases which were involved in the trend. In light of this, in Section VI-B, we use the hot keywords (or phrases) announced by weibo.com and set the experiment duration for from one week before to one week after the peak.

Among the four types of words, 4-character idioms showed the best UP score and high LCP. This is because 4-character idioms are fixed and most of them are from ancient Chinese, there are rarely other phrases which share the same root trigrams with 4-character idioms. This implies that if the words or phrases are stable or fixed then they are not influenced by current events heavily. To test this hypothesis, in Section VI-C, we collect 4-character idioms only from Wiktionary and test them in a 3-month and an 8-month period separately.

### B. Weibo hot keywords

Weibo.com lists 50 hot keywords, or phrases, hourly, daily and weekly. We used the weekly hot keywords of 2 November, 2011 to find out how Connector works on unknown words or phrases with pronounced trends. We picked 7 days before and 7 days after 2 November, 2011 as the frequency vector, so the vector length is 15. As in the previous section, UP is the reasonable results judged by a human (39) divided by the valid root trigrams (43). LCP is the results which match the keywords provided by weibo.com (24) perfectly divided by the valid root trigrams (43). The UP we measured in this experiment is 0.907 (39/43), and the LCP is 0.558 (24/43). These are promising results since some of the phrases are longer than 7 characters, contain several words, and especially have *stop words* in them.

The 7 invalid trigrams are due to the fact that some keywords listed by Weibo are not really hot in terms of what users are actually talking about. For example for 萌物鉴定, the frequency of 萌物鉴 has 13 frequency values equal to 0 and 2 equal to 1. We treated these as invalid root trigrams as described in Section V.

The relatively low LCP value is because we sometimes observe alternate results from the hot keywords provided by

weibo.com. For example, for the hot word 乔布斯情书 (A love letter from Steve Jobs), the result of the connector is 乔布斯传 (Steve Jobs: A Biography). The reason for this is because 乔布斯传 (Steve Jobs: A Biography) has higher frequency and fluctuation than 乔布斯情书 (A love letter from Steve Jobs) during the vector period. We only count them as UP but not LCP. For further error analysis, please refer to Section VII.

We also used the 8-month data to concatenate those keywords and the LCP score was 0 with the UP score less than 0.40. The low LCP and UP score

### C. LCP and UP for known words

To test our hypothesis that fixed words or phrases are not influenced by current events as heavily, we collected all 853 Chinese 4-character idioms listed on Wiktionary [11]. The Connector was executed for two periods, one is about a 3-month period (23 July 2011 to 12 September 2011), the other is an 8-month period (23 July 2011 to 22 March 2011). The results are listed in Table II.

The LCP value needs to be put into context because there are two reasons why it is not as high as the precision of common approaches and the UP score in the same experiment. First, some 4-character idioms have common first root trigram. For example, 一面之交, 一面之识, 一面之词, 一面之雅, and 一面之缘, have the same root 一面之, so Connector only returns the one with the most frequent result: 一面之缘. For this case, we only count 一面之缘 as "matches with Wiktionary," the other four are not counted in LCP, but are counted in UP. Secondly, the words listed in Wiktionary may not necessarily be the common usage of the word. For example, people frequently use 一见如故, as Connector gives, instead of 一见如旧, which is listed in Wiktionary[5].

Though Table II shows that the 8-month period experiment (columns 2 and 4) gives more words that matching the Wiktionary lexicon or are judged correct by a human than the 3-month period (column 3 and 5) does, the LCP and UP values are relatively stable. For those 4-character idioms, they are usually used when people want to express a strong feeling and are seldom involved in pronounced trends. There are more invalid root trigrams in a 3-month period than in an 8-month period experiment, because some trigrams do not appear often enough.

### D. Error Analysis

According to our manual error analysis, the top three errors are the following:

- When test words or phrases contain stop words ( e.g. 的(of), 是(is)) and high frequency common three character words or phrases ( e.g. 会不会 (will)).
- If there are multiple phrases to describe the same thing, or at the time of the experiment period there are multiple memes which contain the same trigrams, then the whole phrase tends to be lengthened, and the results can be a combination of multiple phrases. ( e.g. Both 友谊地久天长 (Our friendship will forever last) and 友谊天长地久 (Our friendship will last forever) are correct, the result becomes 友谊地久天长地久 (Our friendship will last forever last). )

[5]The whole data set and results can be found on the authors' website.

TABLE III
Trigram frequency of 谷歌开发者大会.

| | 26 | 27 | 28 | 29 | 30 | 31 | 01 | 02 | **03** | 04 | 05 | 06 | 07 | 08 |
|---|---|---|---|---|---|---|---|---|---|---|---|---|---|---|
| 谷歌开 | 1 | **68** | 5 | 0 | 1 | 1 | 4 | 0 | 2 | 2 | 4 | 0 | 0 | 3 | 1 |
| 歌开发 | 1 | **68** | 4 | 0 | 1 | 1 | 0 | 0 | 2 | 4 | 0 | 0 | 0 | 0 |
| 开发者 | 49 | **127** | 43 | 46 | 44 | 65 | 50 | 49 | **227** | 129 | 38 | 39 | 65 | 63 |
| 发者大 | 15 | **56** | 5 | 4 | 13 | 10 | 6 | 11 | **166** | 84 | 14 | 11 | 21 | 14 |
| 者大会 | 15 | **56** | 5 | 4 | 14 | 9 | 6 | 11 | **168** | 85 | 14 | 10 | 21 | 13 |

- If the target phrases are composed of multiple words and they have different patterns during the experiment period. For example, the trigram trend of 谷歌开发者大会 (Google's developer conference), one of the hot keywords in our experiment in Section VI-C, are shown in Table III. Trigrams 谷歌开 and 歌开发 only have one peak on 27 October 2011. 开发者, 发者大, and 者大会 have two peaks on 27 October and 3 November 2011 respectively. Connector returns 谷歌开发 (Google develop) incorrectly. We hypothesize that there was another 开发者大会 (Developer conference) held on 3 November 2011.

These three challenges will be the subject of future work.

## VII. Discussion

One of the questions this document focuses on is: does the frequent creation and use of new, unknown words exacerbate the existing problems that natural language processing tasks have with word segmentation, or does it provide an opportunity where the temporal variance in these new words can be leveraged to do a better job of the natural language processing task. In this section, we first use an example to show how a trigram which is not a word can actually help with an information retrieval task. Then, we discuss why we use trigrams.

### A. From trends to stories

To understand what Chinese net users talk about on social media, [we selected 12,891 trigrams which had a chage rate greater than 100 in one day] from 23 July 2011 to 22 March 2012 and tried to concatenate them using an 11 day period (5 days before and 5 days after the increasing date). On average, about 56 trigrams[6] had rates higher than 100 everyday.

Interestingly, we found that the Connector sometimes can tell us the story of the event and what caused a large increment in frequency on that day. Here is an example from our results:

```
100100\_20110804\_万为开:d: gram=万为开,
万为开拓团拍电视,
(Wan made a TV program about the first immigrants)
万为开拓团纪念碑被警,
```

[6]$12,891/229 = 56.3$, where 229 is the number of days we collect the data successfully

| Data period | Valid Root Trigrams | | Match with wiktionary | | Correct (human)[1] | | LCP | | UP | |
|---|---|---|---|---|---|---|---|---|---|---|
| | 8 months | 3 months | 8 months | 3 months | 8 months | 3 months | 8 months | 3 months | 8 months | 3 months |
| $Freq > 99$ | 160 | 122 | 123 | 97 | 6 | 3 | 0.77 | 0.80 | 0.93 | 0.95 |
| $Freq > 29$ | 344 | 279 | 257 | 214 | 12 | 6 | 0.75 | 0.77 | 0.94 | 0.94 |
| $Freq > 4$ | 515 | 484 | 342 | 333 | 23 | 19 | 0.66 | 0.69 | 0.94 | 0.95 |
| Total | 832 | 832 | 364 | 361 | 58 | 56 | 0.43 | 0.43 | 0.92 | 0.90 |

[1] In addition to Matching with Wiktionary.

```
(The statue was splashed ...)
万为开拓团纪念碑被泼红漆,
(The statue was splashed with red oil paint)
万为开拓团纪念碑被5人砸,
(The statue was smashed)
万为开拓团民,
(The first immigrant people)
```

万为开 is a trigram which has no meaning in Chinese. We caught this particular trigram out of the 323 million trigrams in our database because it appeared 100 times more frequently than average on 4 August 2011. After we fed this trigram into Connector and set the connection time from 5 days before to 5 days afterward, we found the phrase: 万为开拓团拍电视 (Wan made a TV program about the first immigrants). It is still not clear enough to tell us why making a TV program created a trend. However, if you read the candidate results, the whole story becomes clear. It tells us that 5 men smashed and splashed red oil paint on the statue of "The First Immigrants."

In this event, there are many trigram words, such as 纪念碑 (statues), 开拓团 (immigrants), 红油漆 (red oil paint) and so on. However, the trigram which had the most significant was, 万为开, a meaningless trigram. In general, these three characters did not appear together before this event. The sudden frequency increase of this trigram from 0 helps our system notice this trigram, which lead us to notice this event. Other trigrams did not increase in rate as much as 万为开 because of this event. This may be because they already exist and thus it is difficult for them to have a precipitous increase in one day. What our preliminary results regarding trend analysis suggest is that the new emergence of the trend of new words can actually be more conspicuous than known words.

From this example, we can see that by re-thinking the order of operations, a natural language processing task can leverage the frequent creation and use of new words, as shown in Figure 3.

Traditional methods need to have a lexicon containing known words or terms to do analysis. After analyzing the trend, they can infer the topics, memes, hot keywords, *etc.*, for only already-known words and terms. However, in Chinese social meida, the unknown words that net users use are an important part of understanding what is being discussed. Our approach is to consider the trends of trigrams first, where we do not need to know the words and the meaning of the words. After analyzing trends, we can get the topics, memes, and keywords for both known words and unknown words and build larger words and phrases from the temporal correlation

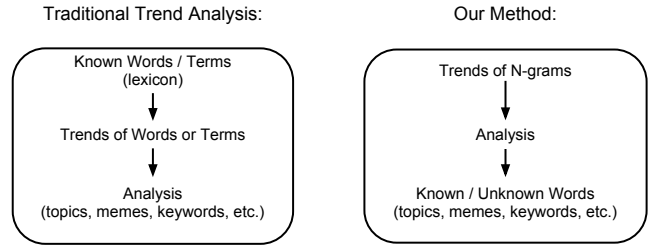

Fig. 3. Change of Order.

of trigrams.

### B. Why trigrams?

As we mentioned in Section III, there can be trends for bigrams, trigrams quadgrams, and so forth. Here we discuss why we focus on trigrams. By combining individual bigrams, trigrams, and the subject of this document which is longer words and phrases connected via trigrams, then we can analyse the trends for words or phrases of any length.

Referring back to the example in Section III, 中华人民共和国 can also be successfully concatenated using bigrams. However, we only investigated the feasibility of using trigrams in this document. We choose trigrams instead of bigrams or quadgrams as a tradeoff of efficiency for accuracy.

Monograms have temporal correlation information, but cannot give us order information of the phrase. For bigrams, the bigram trends of 中华人民共和国 are plotted in Figure 4. Together with Figure 2, we can see that this trend information reflects what we analyzed in Figure 1. We can also see that the trigrams information is more efficient than bigrams in terms of serving as "glue" to hold the larger word together. The reason for this is because 70% of Chinese words are bigrams [12]. Trigrams are a good way to remove the noise of the sub-bi-words in the long word, for example, remove the noise of 人民 from 中华人民共和国.

Next, why not longer grams? Larger grams are computationally expensive. As of 22 March 2012, our dataset has a total of 36,674 monograms, 16,353,985 bigrams, and 323,862,767 trigrams in our database. Figure 5 is the daily increment of distinct monograms, bigrams and trigrams[7]. We can see that the distinct grams are constantly growing, though the growth rate is decreasing slightly. For trigrams, it is notable that there are more than 800 thousand new trigrams added to our database every day.

---

[7] The sudden drop around 26 August 2011 (before day 50) is caused by a system failure for three days around that time. The increase around 12 November 2011 is due to several changes which increased the performance of Collector.

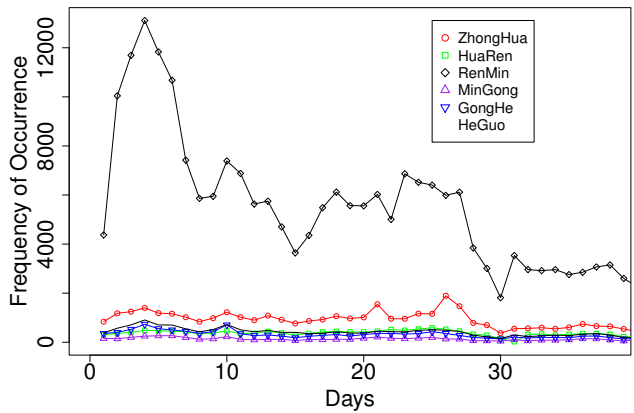

Fig. 4. Bigram trends for 中华人民共和国 on Weibo.

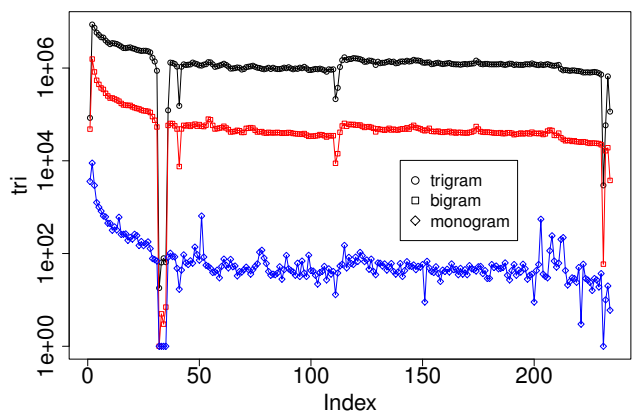

Fig. 5. Distinct n-grams increase rate.

## VIII. RELATED WORK

With regard to natural language processing and social media, our work is distinguished from related work mainly in that related work has not addressed two challenges specific to Internet online corpora: the frequent use of neologisms and the usage of time sequences in the corpus for extra information.

Natural language processing can be a very powerful tool for understanding social aspects of content through data mining. Chao and Xu [16] show some interesting demographics of hate groups on a blog platform. Kwak *et al.* [17] study the dynamics of Twitter and find that re-tweeting causes very fast diffusion of information. Asur *et al.* [18] model the formation, persistence, and decay of trends, and show evidence that, surprisingly, factors such as user activity and number of followers are not strong determinants for the creation of trends. Specific to China, Wei [19] discusses new media research papers published in Chinese-language scholarly journals. Wang *et al.* [20] studies Chinese-language scientist bloggers.

The only two English-language analyses of Weibo that we are aware of are Yu *et al.* [21] and Bamman *et al.* [22]. Yu *et al.* [21] conclude, "People tend to use Sina Weibo to share jokes, images and videos and a significantly large percentage of posts are retweets. The trends that are formed are almost entirely due to the repeated retweets of such media content." [22]

Leskovec *et al.* [15] present a framework for tracking trends on a short time-scale, *i.e.*, tracking "memes." They analyzed information flow between traditional media and blogs during the 2008 U.S. Presidential election. They found that memes, such as "lipstick on a pig," started in the traditional media but moved to blogs within hours. In their mathematical model of memes, only imitation and recency are needed to qualitatively reproduce the observed dynamics of the news cycle.

Unknown word extraction in Chinese is a very challenging problem. Goh [2] presents an approach based on tagging characters and then applying support vector machines and maximum entropy models to the appropriate features. Lu *et al.* [24] propose the use of ant colony optimization for unknown word extraction. Cortez and da Silva [25] have developed an unsupervised approach to information extraction by text segmentation. Search engines also must deal with unknown words, which has been the subject of considerable research [26], [27], [26], [28], [29].

There are several notable entropy-based approaches to word segmentation. Kempe [30] describes a technique for segmenting a corpus into words without information about the language or corpus, and no lexicon or grammar information. This technique works with corpora containing "clearly perceptible" separators such as new lines and spaces. Jin and Tanaka-Ishii [14] propose an unsupervised word segmentation algorithm that is based on the entropy of successive characters at word boundaries. Zhang *et al.* [31] propose a method that is based on a maximum entropy model. Zhikov *et al.* [32] utilize the local predictability of adjacent character sequences, which is analogous to entropy.

## IX. CONCLUSIONS

We propose a pointillism approach for Chinese social media natural language processing. Time is an important aspect of the pointillism approach. Language is viewed as a time sequence of points that represent the grams. The pointillism approach allows us look at a corpus in a different dimension and from a different perspective. In the pointillism approach, trigrams serve as an intermedium to extract trend information.

Our results show that when words and topics do have a meme-like trend, they can be reconstructed from only trigrams. For example, for 4-character idioms that appear at least 99 times in one day in our data, unconstrained precision is 0.93. For longer words and phrases collected from Wiktionary (including neologisms), unconstrained precision is 0.87. We consider these results to be very promising, because they suggest that it is feasible for a machine to reconstruct complex idioms, phrases, and neologisms with good precision without any notion of words. Experiments also tell us that a longer period of frequency of trigrams can be used to connect fixed words, while shorter periods which include the trend can be used to connect phrases, memes and, even find the overall topic. Thus the colorful and baroque uses of language that typify social media in challenging languages such as Chinese may in fact be accessible to machines.

The way we can extract hot words without any aid of dictionary and grammar, is not only useful in languages which do not delimit the words by space, but also help seach query segmentations, dynamic ranking pages, search engine index, *etc.*. for Indo-European languages.